\documentclass[amssymb, aps,twocolumn,groupedaddress, nofootinbib]{revtex4-1}
\pdfoutput=1
\usepackage[cmex10]{amsmath} 
\usepackage[pdftex]{graphicx}
\usepackage{booktabs} 
\usepackage{multirow} 
\newcommand{\figratio}{0.5}

\usepackage[colorlinks=true,linkcolor=blue]{hyperref}
\begin{document}
\title{Mesoscopic analysis of online social networks - The role of negative ties}
\author{Pouya Esmailian, Seyed Ebrahim Abtahi and Mahdi Jalili}
\email[Corresponding authors:]{abtahi@sharif.edu; mjalili@sharif.edu}
\affiliation{Department of Computer Engineering, Sharif University of Technology, Tehran, Iran}

\date{\today}

\begin{abstract}
A class of networks are those with both positive and negative links. In this manuscript, we studied the interplay between positive and negative ties on mesoscopic level of these  networks, i.e., their community structure. A \textit{community} is considered as a tightly interconnected group of actors; therefore, it does not borrow any assumption from balance theory and merely uses the well-known assumption in the community detection literature. We found that if one detects the communities based on only positive relations (by ignoring the negative ones), the majority of negative relations are already placed between the communities. In other words, negative ties do not have a major role in detecting communities of studied signed networks. Moreover, regarding the internal negative ties, we proved that most unbalanced communities are maximally balanced, and hence they cannot be partitioned into \textit{k} nonempty sub-clusters with higher balancedness ($k \ge 2$). Furthermore, we showed that although the mediator triad $++-$ (hostile-mediator-hostile) is underrepresented, it constitutes a considerable portion of triadic relations among communities. Hence, mediator triads should not be ignored by community detection and clustering algorithms. As a result, if one uses a clustering algorithm that operates merely based on social balance, mesoscopic structure of signed networks significantly remains hidden.
\end{abstract}
\keywords{signed social networks, social balance theory, community detection, graph clustering}

\pacs{}

\maketitle

\section{Introduction}
\label{intro}
In the past two decades, there have been increasing interests toward the analysis of complex networks both empirically and theoretically  \cite{albert2002statistical, boccaletti2006complex, dorogovtsev2002evolution}. One of the important research lines is to study networks from the structural point of view, trying to answer, \textit{What do different types of networks look like?} This is an important issue, since it has been shown that many dynamical properties depend on the network structure \cite{barabasi1999emergence, leskovec2008statistical, lancichinetti2010characterizing}. This endeavor is constantly coevolving with the studies on theoretical models of networks trying to describe the observations and further predict new features \cite{strogatz2001exploring,newman2001random}. Most of these works have been carried out due to abundant large scale datasets gathered over the internet. They have attracted a lot of studies mainly to justify the long-standing debates on static and/or dynamic patterns of relations  \cite{backstrom2006group, leskovec2010signed, grabowicz2013distinguishing, grabowicz2012social, onnela2011geographic}.

There are a number of challenges related to signed networks. Discovering the community structure is one of these problems that has been addressed in a number of research works \cite{yang2007community, gomez2009analysis, traag2009community}. Another problem related to these networks is to predict the sign of relations \cite{leskovec2010predicting, tang2012inferring, javari2014cluster}.

Generally speaking, there have been two trends toward the analysis of signed networks. The first trend tries to evaluate the long-standing social balance theory and to deduce some new implications \cite{leskovec2010signed, facchetti2011computing}. The social balance theory has some predictions about the grouping of people based on the analysis of network evolution toward a more balanced structure \cite{zheng2014social}. The second trend, regardless of the balance theory, tries to improve the inference tasks using the negative relations \cite{yang2007community, leskovec2010predicting}. For example, detecting the community of densely interacting individuals is one of the issues studied in such works \cite{traag2009community}. The notion of \textit{community} has been introduced as a meaningful building block of networks \cite{girvan2002community}. Indeed, community structure acts as a bridge between local and global understanding  of network structure \cite{newman2003social, newman2003structurefunc}. In signed networks, grouping the actors has been studied in both community detection and social balance literature \cite{gomez2009analysis, doreian2009partitioning}. In the former, the main objective is expressed as ``dense positive'' and ``negative free'' relations inside groups. In the latter, the objective is explicitly stated as minimizing the number of negative (positive) links inside (between) the groups. These two notions, despite their similarities, have fundamental differences, which are investigated in this work. The main motivation of our work is based on the recent work of Doreian and Mrvar \cite{doreian2009partitioning}. They suggested that the $++-$ relation among groups of individuals is likely to be seen, and thus, it should not be ignored while detecting the mesoscale structure of networks.

As a connection to the above trends, our work starts with the justification of community detection in signed networks and shows that negative relations are not informative enough to improve the detection task. In other words, one can accomplish the task by considering only the positive relations. Our study also deals with the justification of the balance theory in mesoscopic level. Analogous to the local level, this theory states that no matter how (internally balanced) communities are  identified, one must not see (or at least rarely see) the $++-$ triadic relation among them. We found that the observed triads are also underrepresented in mesoscopic level consistent with this theory. However, they form a considerable portion of social relations, which is far more than the corresponding local level, and cannot be simply ignored by clustering algorithms. Therefore, if the social groups are identified based on balance theory, one would miss a considerable amount of distinguishable groups by merging them into one another. Our results shed new light on mesoscale structure of signed networks.
\section{Preliminaries}
\label{sec:problem}
\subsection{Notations}
Throughout the paper, the expressions ``link", ``edge", ``tie", ``relation", and ``interaction" are used interchangeably, unless we explicitly make a note.
A signed graph $G$ is determined using triple $(V, E, \sigma)$. $V$ is the set of nodes, $E$ is the set of edges, which is defined by pair $(v_i, v_j)$ of nodes [$(v_i, v_j) = (v_j, v_i)$ for undirected graph], and $\sigma$ assigns either $+1 $ or $-1$ to each edge. In this work, we consider only undirected signed graphs with values $-1$ and $+1$ for negative and positive relations, respectively.
Having $k$ \textit{nonempty} clusters in a network, let us define the number of inconsistent or frustrated edges as follows:
\begin{equation}
\label{eq:frust}
F_k(G, C) = \sum_{C_i = C_j,i<j} A^-_{ij} + \sum_{C_i \neq C_j,i<j}
A^+_{ij} \mbox{ },
\end{equation}
where $G$ is a signed graph, $C$ determines the cluster of nodes ($C_i$ = cluster to which node $i$ belongs), $k$ is the number of nonempty clusters,  and $A^+_{ij} = 1$ if $\sigma_{ij} = 1$, or $A^-_{ij} = 1$ if $\sigma_{ij} = -1$, or both are zero otherwise. We denote the minimum value of the above function under all possible clusterings as:
\begin{equation}
\label{eq:minfrust}
F_k(G) = \mbox{min}_{C}F_k(G, C),
\end{equation}
where the number of clusters $k$ is a constant value. When $k$ is tunable, one has:
\begin{equation}
\label{eq:minfrust2}
F(G) =  \mbox{min}_{C,k}F_k(G, C).
\end{equation}
In the literature, Eq. (\ref{eq:minfrust}) is often considered as \textit{frustration index} \cite{zaslavsky2010balance}, \textit{true frustration}, or merely \textit{frustration} \cite{facchetti2011computing, iacono2010determining}. However, in this context, the frustration and its minimum are considered separately. For $k = 1$, frustration of a subgraph is equal to the number of negative edges, and thus $F_1(G,C)$ [or equally $F_1(G)$] is used to denote the number of negative edges inside a subgraph. We use $f_k(G,C)$ as the ratio of $F_k(G,C)$ to the edge count $m = |E|$ [similar for $f_k(G)$]. Notations $F_{k,up}(G)$ and $F_{k,low}(G)$ are used for the upper bound of $F_k(G)$ and its lower bound, respectively ($F_{k, low}(G) \leq F_k(G) \leq F_{k,up}(G)$).

Given a specific clustering $C$, we define balancedness of graph $G$ as follows:
\begin{equation}
B_k(G) = 1- f_k(G).
\end{equation}
Generally, we use the term \textit{balanced} when a given subgraph $S$ (i.e., an extracted community) has no negative edges [$B_1(S) = 1$], and \textit{unbalanced} when $B_1(S) < 1$. Note that a graph may have higher balancedness for $k > 1$, which is denoted explicitly throughout the paper.
\subsection{Correlation Clustering problem}
\label{subsec:CC}
In this problem, one seeks to find a clustering of nodes that minimizes inconsistent relations. This is equivalent to minimization of $F_k(G,C)$ considering  $k$ either as a constant value \cite{giotis2006correlation} or a tunable parameter \cite{bansal2004correlation}. We should mention that the maximization of consistent edges has also been considered in the above works, which has different implications from the algorithmic point of view.
\section{Related Works}
\label{sec:related}
In this section, we introduce some of the research lines related to the clustering of signed networks. Conceptually, they could be divided into two categories, where (1) positive links between clusters are penalized, or (2) instead of this punishment, internal density of clusters is rewarded.
\subsection{Structural Balance and Clustering}
\label{subsec:CSB}
\begin{figure}
\centering
\includegraphics[width=\figratio\textwidth]{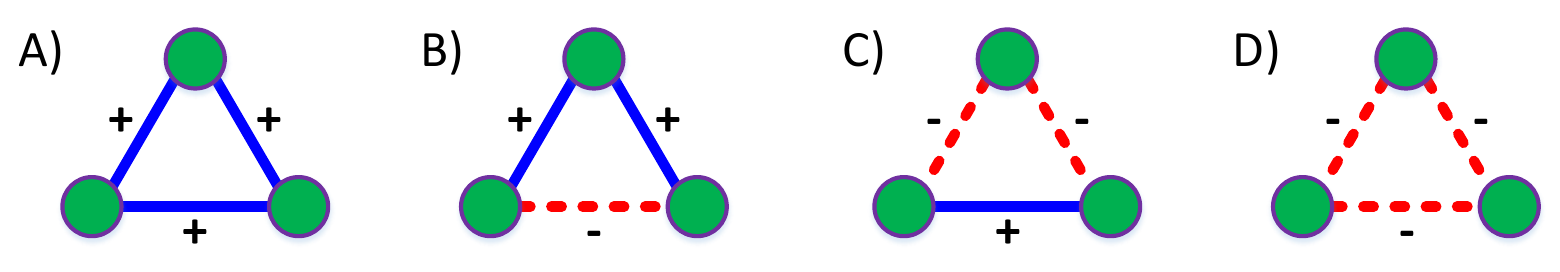} 
\caption{(Color) Different types of triadic signed relations between three actors. In structural balance, triads $A$ and $C$ are balanced, and $B$ and $D$ are unbalanced. In general structural balance, only triad B is unbalanced and the others are balanced.}
\label{fig:triad}
\end{figure}

The origin of structural balance theory is the seminal work of Heider \cite{heider1946attitudes, heider1946attitudes}, which has been further developed as a mathematical framework by Cartwright and Harary \cite{cartwright1956structural}. In the local level, the structural balance theory states that a triadic relation is balanced, if and only if, it has one or three positive ties\footnote{In the structural balance theory \textit{balanced} and \textit{unbalanced} are used only for $k = 2$.}. As shown in Fig. \ref{fig:triad}, triads $A$ and $C$ are balanced, and $B$ and $D$ are unbalanced. In the global level, the structural balance theory states that a graph is structurally balanced (SB), if and only if, it can be partitioned into two clusters with no inconsistent edges (known as \textit{structure theorem}), or equivalently, when every cycle is positive. Inconsistent edges are negative ones inside and positive ones between the clusters. A cycle is positive (or balanced), if and only if, it has an even number of negative links. 

Davis \cite{davis1967clustering} argued that a social network may have multiple hostile groups, implying that triad $D$ is also balanced. In the global level, a graph is $k$-balanced, if and only if, it can be partitioned into $k$-clusters with no inconsistent edge. The term \textit{structural balance} is used for $k = 2$ and \textit{weak-} or \textit{general-structural balance} (GSB) for $k \ge 2$.

To measure the  balancedness of networks, a number of research works have provided some metrics that specify the distance of a graph from GSB \cite{doreian1996partitioning}. In this context, there are two well-known classes of metrics. The first class is based on counting all  unbalanced $l$-cycles (cycles of length $l$), which can only be used for SB. 
The second class is based on counting the minimum number of inconsistent edges under all possible $k$-clusterings  ($=F_k(G)$). In this work, we base our investigations on the second class, and thus, it is briefly discussed in the following. This metric is equal to the minimum number of edges that their deletion (or sign flipping) results in a $k$-balanced graph, which is equivalent to distance of a graph from being $k$-balanced\footnote{This equivalence holds for $k > 2$ with the same proof provided by Zaslavsky \cite{zaslavsky2010balance}.}.

The problem of finding a partition that corresponds to $F_k(G)$ is NP-hard \cite{bansal2004correlation}, even for $k = 2$ \cite{facchetti2011computing}. If we set $k = 2$, the optimal solution is the best two-clustering of a graph where the number of inconsistent edges is equal to the distance of a graph from SB [$=F_2(G)$]. Iacono \textit{et al.} proposed a graph-theoretic approach to approximate $F_2(G)$, which has been originally stated as ``distance from monotonicity" for biological networks \cite{iacono2010determining}; note that monotonicity has the same mathematical implication as SB. The algorithm has been further applied to social networks validating that their distance from SB is significantly lower than those of sign-shuffled counterparts \cite{facchetti2011computing, facchetti2012exploring}. Another achievement of Ref. \cite{iacono2010determining} is a scalable algorithm that calculates a lower bound for $F_2(G)$, which determines, at most, how far is the proposed solution from the optimal value. For $k > 2$, Chiang \textit{et al.} \cite{chiang2012scalable} proposed a scalable $k$-clustering algorithm by transforming an objective function similar to $F_k(G,C)$ (along with some other objectives) into weighted kernel $k$-means. In this paper, we only use the two-clustering algorithm of Iacono \textit{et al.}, together with a theorem that extends our results to $k > 2$.
\subsection{Relaxed Structural Balance and Generalized Block Modeling}
\label{subsec:rsbgbm}
\begin{figure}
\centering
\includegraphics[width=\figratio\textwidth]{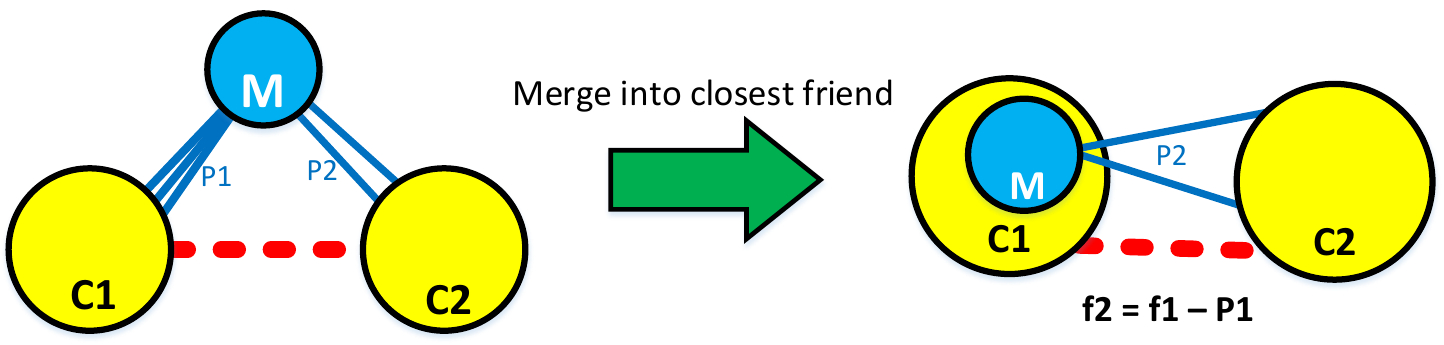} 
\caption{(Color) ``$++-$" relation (hostile-mediator-hostile) between three clusters. The value of $F(G)$ is reduced by $P1$, if mediator cluster $M$ is merged into the closest friend $C1$.}
\label{fig:mediator}
\end{figure}

In contrast to the implications of GSB, Doreian and Mrvar \cite{doreian2009partitioning} argued that real-world networks are not completely balanced. Accordingly, it has been shown that in online social networks with 17-23\% negative ties, at least 7-14\% of edges are inconsistent with SB \cite{facchetti2011computing}. As a result, Doreian and Mrvar proposed the relaxed structural balance (RSB) theory stating that positive interactions between two clusters are also valid. This relaxation is mainly due to intermediary processes in social networks, implying that it is likely to find a mediator group with positive relations toward two hostile groups [Fig. \ref{fig:mediator}(left)].

\begin{figure}
\centering
\includegraphics[width=\figratio\textwidth]{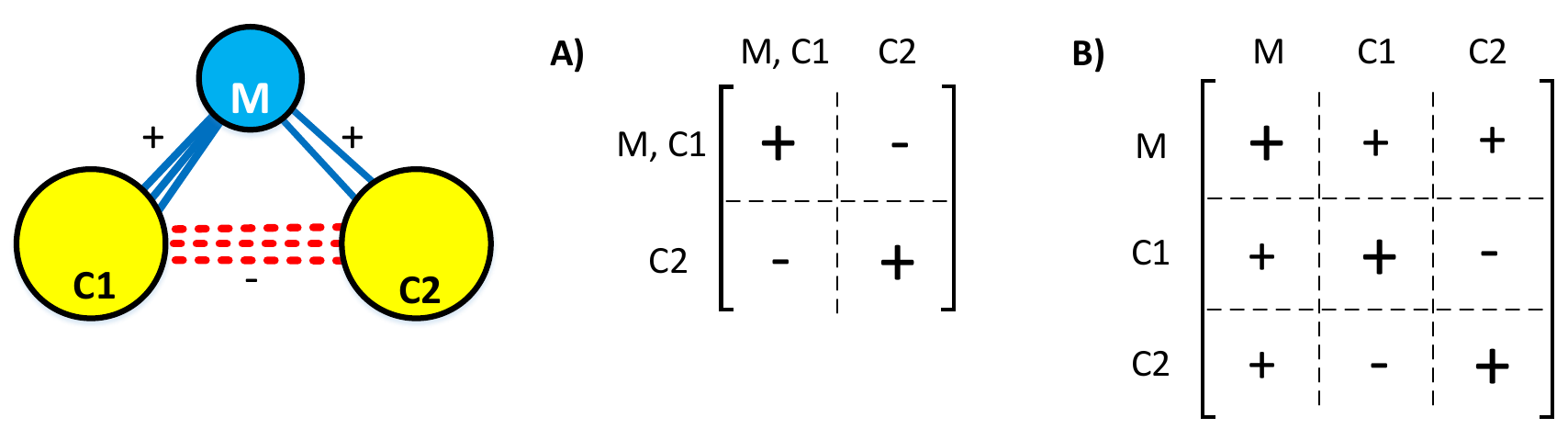} 
\caption{(Color) (A) Result of fitting a GB model based on general structural balance, which merges $C1$ and $M$ to get a lower frustration. (B) Result of fitting a GB model based on relaxed structural balance, which allows off-diagonal positive blocks similar to the output of signed community detection methods, if one restricts the clusters to be internally cohesive.}
\label{fig:gsbrsb}
\end{figure}

Based on GSB, positive edges between clusters are punished.  Hence, as depicted in Fig. \ref{fig:mediator}(right), a mediator cluster is merged into one of the hostile clusters with which it has more positive connection $P1$, decreasing the frustration from $F(G)$ to  $F(G) - P1$. Accordingly, Doreian and Mrvar argued that, based on GSB, blocks (cluster-cluster relations) of positive ties are not allowed in off-diagonal positions of the relation matrix (as shown in Fig. \ref{fig:gsbrsb}). In Fig. \ref{fig:gsbrsb}, one can see the result of fitting a generalized block model (GB model) \cite{doreian2005generalized} on hostile-mediator-hostile triad (mediator triad for short) based on GSB and RSB. However, in order to fit the relaxed model to data, $F_k(G,C)$ is still used in Ref. \cite{doreian2009partitioning} as the objective function. This means one should take care of each off-diagonal positive block \textit{a priori} to refrain the optimization process, which tries to minimize $F_k(G,C)$, from merging mediator clusters into hostile parties.

\subsection{Community Detection in Signed Networks}
\label{subsec:signcom}
In the community detection literature, mainly started after the seminal work of Girvan and Newman \cite{girvan2002community}, there has been a different perspective toward the group identification.  As the main assumption, a community is a group of nodes that have more connections inside than to the rest of the network. This intuition has been the basis of almost all community detection algorithms \cite{fortunato2010community, rosvall2008maps}. Regarding this, \textit{modularity} function has been introduced that gives a better score to a cluster with denser relations than a null model \cite{newman2004finding}. The formulation of modularity allows for straightforward extension to signed networks \cite{gomez2009analysis}. The intuition is that the group of nodes should have more (less) positive (negative) intra-density relative to the null model. This intuition could be formulated by subtracting the modularity score of negative subgraph $G^-$ from positive subgraph $G^+$ as follow:
\begin{equation}
Q(G,C) = \alpha Q(G^+,C) - (1-\alpha) Q(G^-,C),
\end{equation}
where $0 \le \alpha \le 1$  is the relative importance of positive ties compared to negative ones. A similar work has been carried out for Hamiltonian function of Potts model, which borrows the idea of modularity by incorporating an arbitrary null model with a resolution parameter  \cite{traag2009community}. As a summary, these methods reward (punish) the density of intra-positive (intra-negative) relations and punish (reward) their sparseness relative to the null model. Another work extends the community detection based on random walks \cite{yang2007community}, with the intuition that a random walker is more likely to be trapped inside a community. In the main step of the algorithm, the nodes are sorted according to their distance from a sink node. This step ignores the information of negative ties, which are only used as a cut criterion on the sorted list. 

In all of these extensions, there is no explicit punishment strategy for positive edges between the communities, which makes them applicable to nonsigned (or sparsely signed) networks. As a connection to generalized block modeling, these algorithms work with dense diagonal blocks (relations inside a cluster) and sparse off-diagonal blocks (relations between clusters) for positive relations and the reverse for the negative ones. Fig. \ref{fig:gsbrsb} shows a toy example where signed community detection and RSB produce the same clustering that is different from the one produced by GSB.
\section{Community vs Cluster}
\label{sec:comcls}
\begin{figure}
\centering
\includegraphics[width=\figratio\textwidth]{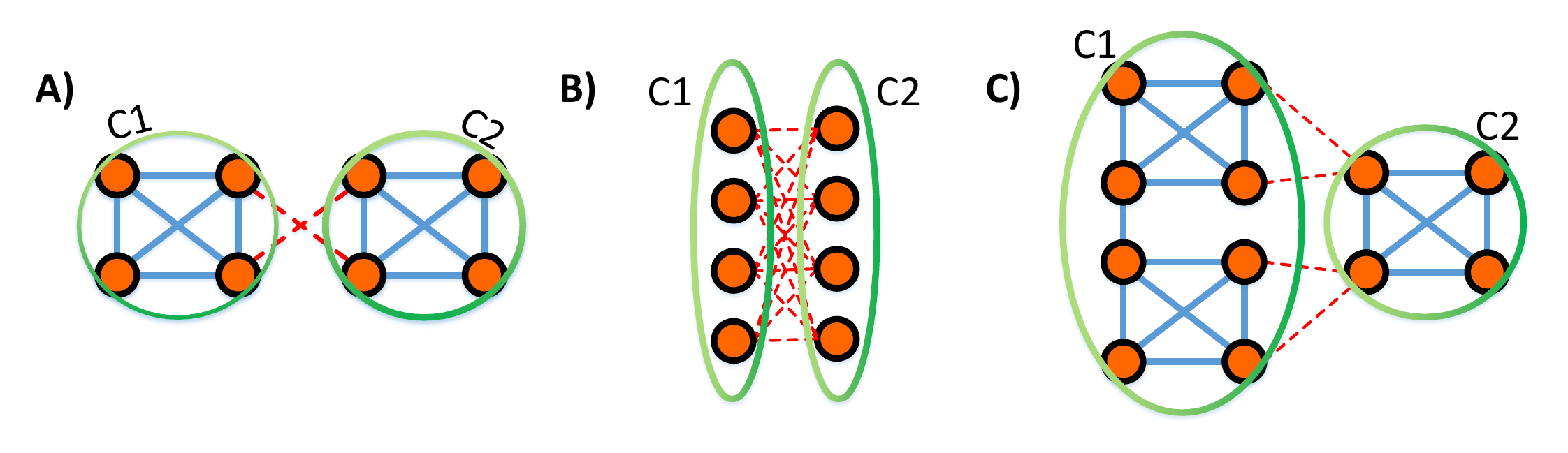} 
\caption{(Color) All three two-clusterings result in $F_2(G,C) = 0$. $A)$ \textit{Cluster} and \textit{community} are consistent with each other. $B)$ Clusters are not \textit{communities}. $C)$ Two distinguishable communities get clustered into $C1$.}
\label{fig:ClusCom}
\end{figure}

In this section, we try to pinpoint some implications of the algorithms that try to optimize $F_k(G,C)$ against those that are frequently used for community detection. The main differences are illustrated in Fig. \ref{fig:ClusCom}. In all cases, detected clusters result in $F_2(G) = 0$  as the optimal solution. In Fig. \ref{fig:ClusCom}(A), the output of clustering algorithms is consistent with the notion of \textit{community}, which is also produced by relaxed GB modeling. However, in Fig. \ref{fig:ClusCom}(B), members of each cluster are disconnected, and thus, despite their similar role in the network, they cannot be considered as a social group of interacting individuals. Also, Fig. \ref{fig:ClusCom}(C) two distinguishable communities are grouped together missing an obvious pattern of relations. These cases [Figs. \ref{fig:ClusCom}(B)-\ref{fig:ClusCom}(C)], as well as the mediator triad are the main shortcomings of clustering algorithms in social networks. Trying to relate these two notions, a community is a cluster of nodes that is internally well-connected. One of the aims of this work is to investigate the frequency of such cases in real networks. We show that the case as shown in Fig. \ref{fig:ClusCom}(C)), as well as mediator triads, are frequent enough that cannot be ignored when one deals with large-scale social networks.
\section{Methods}
\label{sec:method}
\subsection{Extracting Communities}
\label{subsec:excom}
We want to extract groups of densely interconnected nodes that are consistent with the notion of \textit{community}. To this end, we use InfoMap \cite{rosvall2008maps, rosvall2011multilevel}, which is proved to be one of the most accurate community detectors \cite{ahn2010link}. It confidently extracts the communities from large-scale networks that have heterogeneous group sizes \cite{lancichinetti2009community, lancichinetti2010characterizing}. We used the open source code provided in Ref. \cite{website:rosvall} utilizing the hierarchical mode that refines a few big communities into smaller ones, and leaves other communities intact. As studied by Lancichinetti and Fortunato \cite{lancichinetti2009community}, if a group of nodes is well-separated from the environment, it could be accurately detected by InfoMap. However, if the density among some groups passes a threshold, InfoMap mistakenly considers them as a single community. Indeed, this problem happens for all methods that merely consider the structure of a network. In the case of InfoMap, we are confident about the internal density of detected communities relative to the environment \cite{kawamoto2014map}, and as the only problem, there might be more than one group in a single community. Nevertheless, as we illustrate in the results, this problem does not significantly affect the outcome, and the conclusion drawn from the results remains valid.
\subsection{Computing the distance from Structural Balance}
\label{subsec:distance}
As we mentioned in Sec. \ref{subsec:CSB}, for graph $G$, the distance from SB is $F_2(G)$, which is equal to the minimum inconsistent edges under all two-clusterings. Although the computation of this value is NP-hard, the scalable algorithm of Iacono \textit{et al.} \cite{iacono2010determining} outputs a two-clustering, which is an upper bound for $F_2(G)$, as well as a lower bound for $F_2(G)$. Thus, we always know, at most, how far is the sub-optimal solution from an optimal one. Same as Ref. \cite{iacono2010determining}, quantity $F_{2,low}(G)/F_{2,up}(G)$ is used to measure the precision of a solution. Considering the following inequality:
\begin{equation}
\frac{F_{2,low}(G)}{F_{2,up}(G)} \le \frac{F_2(G)}{F_{2,up}(G)} \le 1, 
\end{equation}
if $F_{2,up}(G) = F_{2,low}(G)$, an optimal solution is found. We propose a theorem that generalizes our results to $k$-clustering for $k > 2$:
\newtheorem{theorem}{Theorem}
\begin{theorem}
\label{theo:1}
If $F_1(G) \le F_2(G)$, then $F_1(G) \le F_k(G)$ for every $k > 2$; where every cluster is nonempty\footnote{Note that $F_k(G)$ is defined over $k$ \textit{non-empty} clusters; otherwise, trivially $F_k(G) \le F_1(G)$ for every $k$.}.
\end{theorem}

\textit{Proof.} The proof is through induction. Suppose the theorem holds for $k = 2, ..., k - 1$ and there exists a $k$-clustering that results in $F_k(G) < F_1(G)$. Consider $A_{C,C'}^+$ ($A_{C,C'}^-$) as the number of positive (negative) links between clusters $C$ and $C'$. In such $k$-clustering, links from each $C$ toward every other $C'$ must satisfy $A_{C,C'}^- \ge A_{C,C'}^+$. Otherwise, by merging $C$ into such $C'$, inequality $F_{k - 1}(G) < F_k(G)$ is reached, implying the $F_k(G) < F_1(G) < F_k(G)$ contradiction. With this restriction, if there is no $C1$  satisfying $A_{C1,C'}^- > A_{C1,C'}^+$ for some $C'$, the $F_k(G)  = F_1(G)$ contradiction is reached via merging all clusters into one cluster. Otherwise, we select such $C_1$ and merge all other clusters into $C_2$. Consequently, we find a two-clustering that satisfies $A_{C_1,C_2}^- > A_{C_1,C_2}^+$, and therefore, results in the $F_2(G) < F_1(G) \le F_2(G)$ contradiction. The proof is complete with this. $\square$

Reminding that $F_1(G)$ is the number of negative edges in  graph $G$, theorem \ref{theo:1} states that if an optimal two-clustering has worse frustration than the one-clustering, then every $k$-clustering is also worse than the one-clustering, and thus, it is maximally balanced. As a result, if we get $F_1(G) \le F_{2,low}(G)$ from Iacono algorithm, which signifies $F_1(G) \le F_2(G)$, we conclude that inconsistent edges in $G$ cannot be reduced (or equally, balancedness cannot be increased) via $k$-clustering for $k \ge 2$. Thus, $G$ is optimally clustered into one cluster.
\section{Datasets}
\label{sec:dataset}
We used two widely studied online signed networks known as \textit{Slashdot} and \textit{Epinions} \cite{leskovec2010predicting},  which have been frequently used as benchmarks for studying signed social relations \footnote{All datasets are publicly available at \url{http://snap.stanford.edu}. For more detailed statistics refer to \url{http://konect.uni-koblenz.de/}}. These datasets have special characteristics that make them suitable for the analysis of social relations. For example, all the links have been explicitly established by the users, either positive (for friendship or trust) or negative (for enmity or distrust). Hence, the links neither have been inferred indirectly nor been asked from a person, which may introduce biasedness into data.
\subsection{Data preprocessing}
\label{sec:datapre}
We performed some preprocessings on the datasets preparing them for our purpose:
\begin{enumerate} \itemsep4pt \parskip0pt \parsep0pt
\item In order to get an undirected network, reciprocal links with inconsistent signs were omitted, and the remaining links were considered as undirected [inconsistent relations were 0.7\% (0.4\%) of relations in Epinions (Slashdot)].
\item Only the largest connected component of each network was considered (90\% of nodes in Epinions and nearly 100\% of nodes in Slashdot).
\item Nodes incident to zero positive edges were removed as they, trivially, belong to an isolated cluster.
\item After detecting communities, we kept only those of size 3 to 2000 with all connections between them; the reason is provided in the following.
\end{enumerate}
Table \ref{tab:basic} summarizes the properties of networks after the above operations.
The community size is lower bounded to 3, which is the minimum trivial group size. We did not consider mega-scale communities of size larger than 2000 (4 out of 10k communities that have 3k, 5k, 8k, and 10k nodes), because either they are a composition of many highly interconnected sub-communities, or they have no community structure at all. As a result, they cannot be counted as reliable social communities. Indeed, the size of these communities is significantly far from 150, which is the expected upper-limit for human community \cite{leskovec2008statistical}. In addition, the significantly high $ f_1(G)$ of the largest community in each network fortifies this conjecture. Nonetheless, we further analyze them along with the other unbalanced communities in Sec. \ref{sec:imbal}, and found similar results for the role of negative edges that lie inside them.
\begin{table}
\centering
\caption{Basic statistics of datasets after preprocessing. Average members is the mean of community sizes.}
\label{tab:basic}

\begin{tabular}{|c|c|c|c|c|c|}
\hline
\toprule
      & Node  & Edge  & \parbox[c]{2cm}{Negative Edge} & Community &  \parbox[c]{1.5cm}{Average  members} \\
\midrule
\hline
Slashdot & 68409 & 327490 & 69682 (21.27\%) & 4598  & 14.88 \\
\hline
Epinions & 76653 & 220932 & 13921 (6.30\%) & 6032  & 12.71 \\
\bottomrule
\hline
\end{tabular}
\end{table}
\section{Interplay between Dense Positive and Negative Ties}
\label{sec:interplay}
We discussed that the community detection problem in signed networks is to find groups of densely connected positive ties that are as balanced as possible. First, one needs to get an image of interplay between dense positive ties and those with negative sign. To this end, we first detect the communities from positive subgraph of preprocessed networks using InfoMap. In other words, we exclude the negative subgraph and ignore the information given by negative ties. Next, we bring the negative ties back to the network, noting that the communities have been detected beforehand. Considering only the communities of size 3 to 2000 and the connections between them, we find that more than 98\% of communities are completely balanced, meaning they contain no negative relations (lower bounding the size to 10 also gives similar results). Knowing that unbalanced communities are mostly the bigger ones, we also find that more than 98\% of negative ties lie between communities (see Table \ref{tab:info} for more detailed statistics). These results are interesting, since we based our community detection merely on positive ties and ignored the negative ones. One immediate conclusion is that negative ties naturally lie between densely connected positive ties, and thus, both objectives ``densely connected positive ties inside cluster" and ``negative ties between clusters" could be reasonably satisfied without considering the latter. In other words, positive ties have the major role in detecting the community structure in signed networks, whereas negative ties have a minor effect. These results, somehow, legitimize the idea behind FEC algorithm \cite{yang2007community}, which scores the nodes regardless of negative ties; however, this may not be the case for other types of networks. This observation is consistent with the findings of Leskovec \textit{et al.} that are based on the analysis of triads \cite{leskovec2010signed}. In particular, they concluded that negative ties tend to act like bridges in signed social networks. Nevertheless, due to relatively low amount of negative ties (around 21\% in Slashdot and 6\% in Epinions after preprocessing), it may not be a significant observation and could be highly probable in random counterparts of observed networks; this issue is investigated in Sec. \ref{subsec:sig}. Moreover, unbalanced communities, which are mostly the big ones, are analyzed separately in Sec. \ref{sec:imbal} to investigate the role of negative ties inside them.
\begin{table}
\centering
\caption{Community statistics of studied online social networks. Solved negative ties are links that lie between communities. Average frustration is the mean of $f_1(G)$ over communities.}
\label{tab:info}
    \begin{tabular}{|c|c|c|c|c|}
    \hline
    \toprule
    \multirow{2}[4]{*}{} & \multicolumn{2}{c|}{Slashdot} & \multicolumn{2}{c|}{Epinions} \\ \cline{2-5}
    \midrule
          & Count & Percentage & Count & Percentage \\
	\hline
    Balanced Communities & 4543  & 98.80\% & 5952  & 98.67\% \\
    \hline
    Solved Negative Ties & 68794 & 98.73\% & 13737 & 98.67\% \\
    \hline
    Average Frustration & \multicolumn{2}{c|}{0.08\%} & \multicolumn{2}{c|}{0.05\%} \\
    \bottomrule
    \hline
    \end{tabular}
\end{table}
\subsection{How significant are the observed statistics?}
\label{subsec:sig}
\begin{figure*}
\centering
\includegraphics[width=0.8\textwidth]{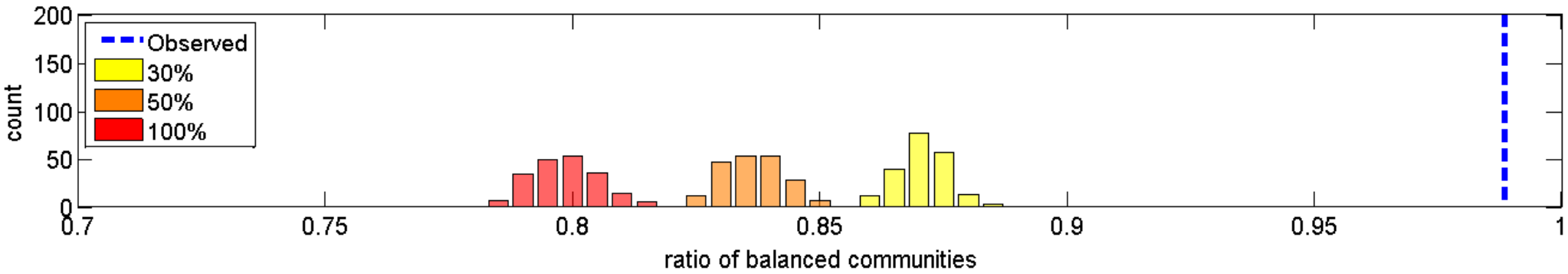} 
\includegraphics[width=0.8\textwidth]{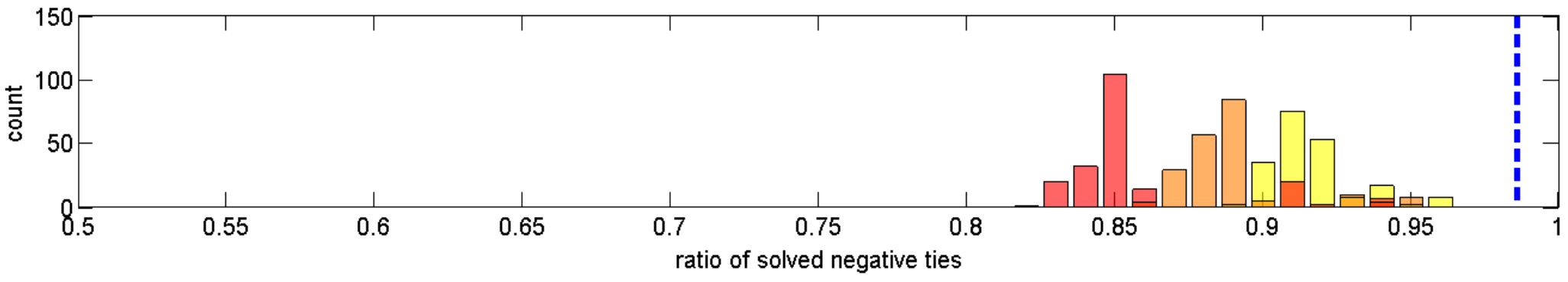} 
\includegraphics[width=0.8\textwidth]{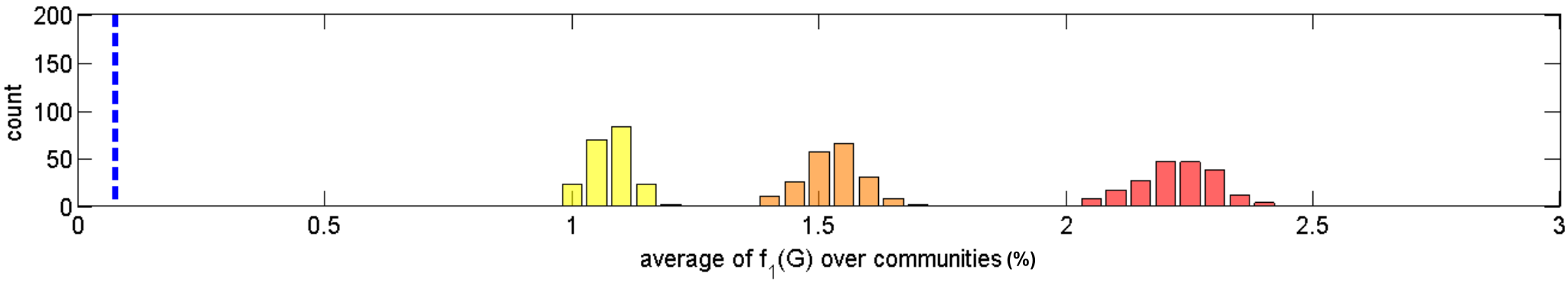} 
\caption{(Color) Observed statistics (dashed blue line) in Slashdot dataset as compared to those of 200 realizations from $M_{r}(G)$ for different values of $r$.  In all cases, $p$-value is less than 0.001.}
\label{fig:slash_pert}
\end{figure*}

\begin{figure*}
\centering
\includegraphics[width=0.8\textwidth]{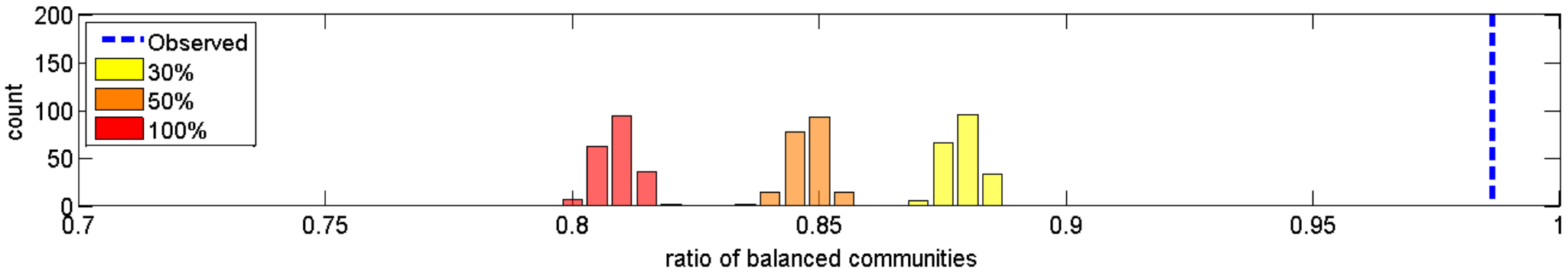} 
\includegraphics[width=0.8\textwidth]{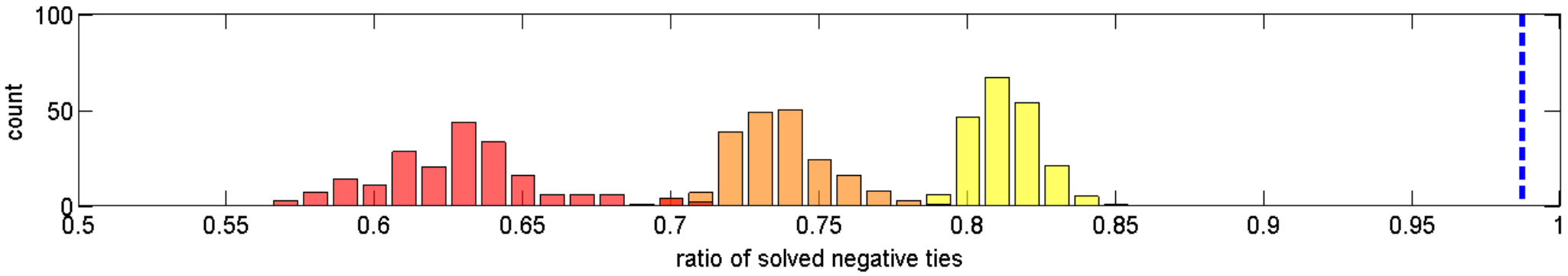} 
\includegraphics[width=0.8\textwidth]{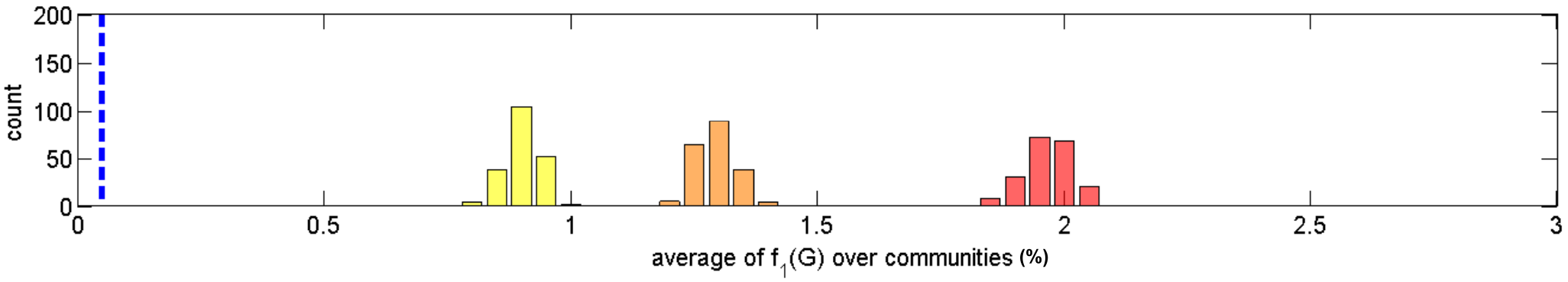} 
\caption{(Color) Observed statistics (dashed blue line) in Epinions dataset as compared to those of 200 realizations from $M_{r}(G)$ for different values of $r$.  In all cases, $p$-value is less than 0.001.}
\label{fig:epin_pert}
\end{figure*}

In order to show the significance of observed statistics in signed networks, first we should define a proper null model to estimate the probability of desired statistics being as extreme as the observed ones. If the estimated probability is small enough, one can conclude that the observed statistics cannot be due to the chance and depend on the characteristics that have been randomized in the null model. We want to show that this significance is due to the particular position of negative edges between dense positive regions. In order to achieve this goal, we proposed  null model $M_{r}(G)$ that is sampled by perturbing $r$ percent of negative links on graph $G$ while keeping the structure of the network fixed. In particular, for a given graph $G$, we select $r$ percent of negative edges uniformly at random and flip their sign to positive, then randomly select the same amount from positive edges and flip their sign to negative. In this setting, the relative number of negative edges and the structure of networks generated from $M_r(G)$ resemble the observed one, and only, the position of $r$ percent of negative edges is randomly shuffled. This null model has been previously used in Refs. \cite{facchetti2011computing} and \cite{leskovec2010signed} for $r = 100$. The complete procedure of acquiring a sample statistic from $M_{r}(G)$ is as follows:
\begin{enumerate} \itemsep4pt \parskip0pt \parsep0pt
\item Perturb $r$ percent of negative ties.
\item Apply InfoMap on the positive subgraph.
\item Bring the negative ties back.
\item Measure the desired statistic.
\end{enumerate}
In Figs. \ref{fig:slash_pert} and \ref{fig:epin_pert}, the observed statistics are plotted in dashed blue line along with 200 realizations of $M_{r}(G)$ for $r =$ 30, 50, and 100 on Slashdot and Epinions networks, respectively. Due to the sufficient number of samples (far more than 30), z-test could be confidently used for computing the $p$-value. This probability is very small ($p \ll 0.001$) for all statistics in both networks compared to the null models with $r =$ 30\%, 50\%, and 100\%. Consequently, it could be concluded that the ratio of balanced communities and solved negative ties are significantly higher, and average frustration [average of $f_1(G)$ over communities] is significantly lower than being created by chance. Furthermore, since we merely flipped the sign of negative ties, observed phenomenon is due to the topological position of negative ties implying that \textit{negative ties almost entirely lie between dense positive ties}.
\subsection {Are considerable parts of the networks isolated from negative ties?}
\label{subsec:isolated}
One of the plausible causes for the observed statistics would be the isolation of major parts of the networks from negative ties, which can lead to numerous balanced communities. However, over 91\% of balanced communities in both networks are incident to at least one negative edge, and the average percentage of external negative ties is around 28\% for balanced communities. Although, this is 5-10\% lower than that of unbalanced communities, it is sufficient enough to reject the \textit{major parts of networks are free from negative ties} hypothesis.
\subsection{Effect of InfoMap on the observed statistics}
\label{subsec:subsubeff1}
As we discussed in Sec. \ref{subsec:excom}, InfoMap may merge highly interconnected communities into each other. First, 98\% of communities are balanced and more than 85\% of nodes in unbalanced communities are incident to only positive edges, thus by further partitioning these communities, the 98\% statistics, if not increasing, would not considerably decrease. Second, by separating mistakenly-glued communities, the number of inter-negative (or -positive) ties does not decrease, which is a trivial case. Finally, the average frustration also follows the first case, and  thus, it does not considerably increase. Consequently, the three main reported statistics are valid and do not considerably depend on InfoMap algorithm.

\section{Analysis of unbalanced communities}
\label{sec:imbal}
So far, we have shown that over 98\% of communities are balanced, which means they have no internal negative ties. On the other hand, unbalanced communities, which are less than 2\% of total communities, are mostly the bigger ones in terms of the number of nodes and ties, and thus, they should be analyzed to find the role of negative ties lying inside them. 

Table \ref{tab:unbalcom} shows the major unbalanced communities including those of size larger than 2000. Considering only communities smaller than 2000 nodes, they have far less negative ties compared to positive ones [$f_1(G)$]. However, this does not mean negative ties are useless from signed community detection or $k$-clustering point of view. In order to measure the usefulness of these negative edges for extraction of balanced clusters, we propose a simple information-theoretic measure that is based on structural balance theory.
\begin{table}
\centering
\caption{Result of applying two-clustering algorithm of Iacono on  unbalanced communities of InfoMap. Percentages are based on total number of unbalanced communities.}
\label{tab:zeroinfo}
\begin{tabular}{|c|c|c|c|}
\hline
\toprule
      & \parbox[c|]{2.5cm}{Unbalanced Communities} & \parbox[c|]{2cm}{Optimal $2$-clustering} & \parbox[c|]{2.5cm}{Zero Information} \\
      \hline
\midrule
Slashdot & 56    & 55 (98\%) & 48 (86\%) \\
\hline
Epinions & 83    & 81 (98\%) & 77 (93\%) \\
\bottomrule
\hline
\end{tabular}
\end{table}
\begin{table*}
\centering
\caption{Detailed statistics of the five largest unbalanced communities in Slashdot and Epinions networks. Bold communities have sizes larger than 2000, which were excluded from preprocessed networks (C1 in Slashdot, C1-2-4 in Epinions). Optimal two-clustering is achieved for those communities that have $\frac{F_{2,low}(G)}{F_{2,up}(G)}=1$.}
\label{tab:unbalcom}
\begin{tabular}{|c|c|c|c|c|c|c|c|c|}
\hline
\toprule
\multicolumn{9}{|c|}{\textbf{Slashdot}} \\
\hline
\midrule
\multicolumn{1}{|c|}{} & Edge & Node & $f_1(C_i)$ & $f_{2,up}(C_i)$ & $f_{2,low}(C_i)$ & $\frac{F_{2,low}(C_i)}{F_{2,up}(C_i)}$   & Larger Partition & $I(G)$ \\
\hline
\textbf{C1} & 33552 & 5378  & 8.57\% & 7.97\% & 7.94\% & 0.9955 & 99.00\% & 0.015 \\
\hline
C2    & 16156 & 1705  & 1.46\% & 1.40\% & 1.40\%  & 1 & 99.65\% & 0.005 \\
\hline
C3    & 6485  & 1204  & 5.97\% & 4.61\% & 4.61\%  & 1 & 98.50\% & 0.022 \\
\hline
C4    & 4700  & 1320  & 2.55\% & 2.49\% & 2.49\%  & 1 & 99.92\% & 0.001 \\
\hline
C5    & 3896  & 1095  & 0.49\% & 0.49\%  & 0.49\% & 1 & 100\% & 0 \\
\hline
\multicolumn{9}{|c|}{\textbf{Epinions}} \\
\hline
\multicolumn{1}{|c|}{} & Edge & Node & $f_1(C_i)$ & $f_{2,up}(C_i)$ & $f_{2,low}(C_i)$ & $\frac{F_{2,low}(C_i)}{F_{2,up}(C_i)}$   & Larger Partition & $I(G)$ \\
\hline
\textbf{C1} & 114989 & 10568 & 11.40\% & 8.54\% & 8.50\% & 0.9949 & 97.56\% & 0.036 \\
\hline
\textbf{C2} & 66314 & 8312  & 5.93\% & 5.15\% & 5.14\% & 0.9988 & 99.01\% & 0.014 \\
\hline
C3 &  36931 & 1033  & 0.05\% & 0.05\%  & 0.05\%  & 1 & 100\% & 0 \\
\hline
\textbf{C4} & 12366 & 2981  & 0.31\% & 0.31\%  & 0.31\%  & 1 & 100\% & 0 \\
\hline
C5 & 6346  & 1043  & 0.77\% & 0.77\%  & 0.77\%  & 1 & 100\% & 0 \\
\bottomrule
\hline
\end{tabular}
\end{table*}
\subsection{How informative are negative ties?}
\label{subsec:info}
\begin{figure}
\centering
\includegraphics[width=\figratio\textwidth]{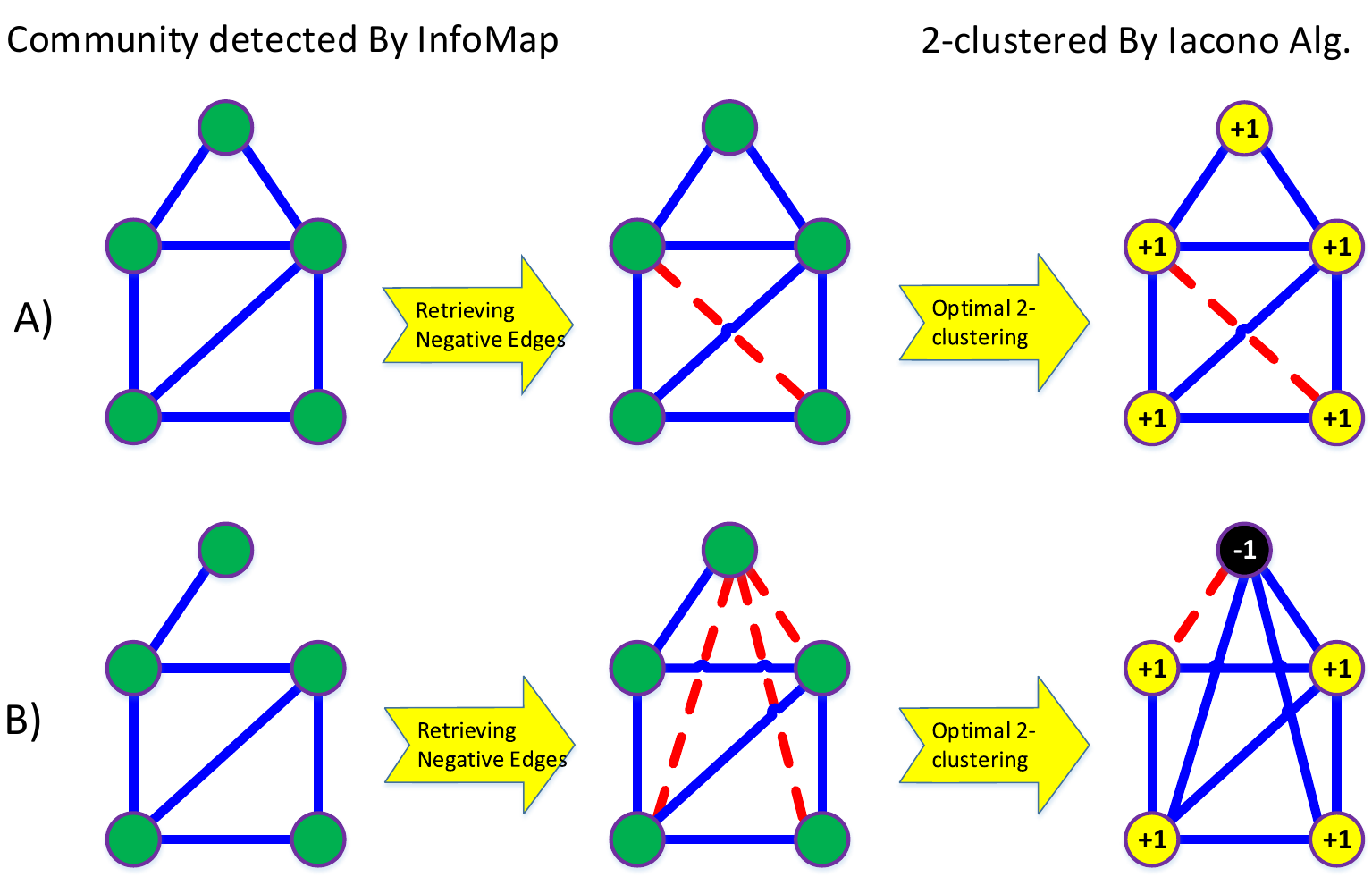} 
\caption{(Color) $(A)$ Information of negative edge is equal to $-log_2(1) = 0$, which also means $F_1(G) \le F_2(G)$. $(B)$ Information of negative edges is equal to $-log_2(\frac{4}{5}) = 0.322$.}
\label{fig:info_example}
\end{figure}

From the structural balance perspective, at one extreme, negative edges inside community $S$ are in the most informative position, if there exists an optimal two-clustering for $S$ that results in two equally-sized clusters. From another point of view, one can argue that by ignoring negative ties, two maximally balanced, equally sized clusters are mistakenly considered as a single community. At the other extreme, negative ties are in the least informative position, if the minimum number of inconsistent ties is achieved by putting $S$ into one cluster. In other words, the same community is achieved with or without considering the negative edges. With this intuition in mind, we define the information of negative ties as follows:
\begin{equation}
I(G) = -log_{2}(\mbox{ratio of nodes in larger partition}),
\end{equation}
where \textit{larger partition} is obtained from an optimal two-clustering, which has $F_2(G)$ inconsistent ties. Whenever $F_1(G) \le F_2(G)$ (no improvement upon one-clustering), the two-clustering is set to one-clustering. This measure is illustrated in Fig. \ref{fig:info_example} for some toy graphs. 

As shown in Table \ref{tab:zeroinfo} and detailed in Table \ref{tab:unbalcom}, for the five largest unbalanced communities, by applying Iacono algorithm we find an optimal two-clustering for 136 out of 139 unbalanced communities. The information of 125 unbalanced communities is zero, for which an optimal solution has been found. Using theorem \ref{theo:1}, this means unbalanced communities cannot reach a higher balancedness by being further partitioned into $k$ nonempty clusters, and thus, they are inseparable from GSB point of view. Moreover, for those with $I(G) > 0$ (11 of 139), separated clusters are relatively very small, and also they are, internally, highly disconnected. This indicates there is no significant sub-community that can be separated from the original one.

In this section, we showed that negative ties inside unbalanced communities are not effectively informative from community detection or $k$-clustering point of view. However, the established results are from two widely studied social networks and should be further investigated on other large-scale ones.
\section{Relation between InfoMap and Signed Modularity}
\label{sec:simod}
We argued in Sec. \ref{subsec:signcom} that the objective of signed modularity is in line with the community detection literature. Therefore, in the absence of negative ties, the goal is the same for both InfoMap and modularity. However, there still remain some major problems. First, as our experiments show, non-signed modularity ($\alpha = 1$) is incapable of distinguishing communities effectively. In particular, the output is mostly made of a few mega-scale communities of size 2k to 20k, which cannot be reliably considered as \textit{single} community (especially for Epinions). Second, modularity suffers from the well-known resolution limit, stating that it is expected to have trivially distinct communities being grouped even in medium-scale networks \cite{fortunato2007resolution}.

In agreement with our results, by sliding the parameter $\alpha$ from 1 to 0.5  (increasing the effect of negative ties with respect to the density of positive ties), the percentage of solved negative ties (those placed between communities) remains almost constant around 88\% in Slashdot. That is, signed modularity is incapable of effectively reducing the internal negative ties. However, in the case of Epinions, this percentage goes from 71\% to 82\%, which means 11\% of negative ties manage to break the communities apart and get placed between them. Nonetheless, we argue that these negative ties are not informative in a general sense. The motivation is that the less an algorithm is capable of distinguishing the communities, the more information it gets from negative ties. In other words, the usefulness of negative ties depends on the performance of a detector. This is intuitively correct, since one can build detector $D$  as:
\begin{equation}
D=\left\{\begin{matrix}
\mbox{One-clustering} & \frac{m^-}{m} < x\\ 
\mbox{Signed modularity} & \mbox{o.w.}
\end{matrix}\right.,
\end{equation}
which puts all the nodes in one community until a certain ratio of negative ties is reached ($x$), and uses the signed modularity afterward. In this case, even if negative ties are truly placed between dense positive ones, they are still useful for detector $D$, since by exceeding $x$, the percentage of solved negative ties increases. This example suggests that the presence of mega-scale communities along with the resolution limit of modularity refrains us from saying that \textit{negative ties are informative} for Epinions. However, if one can find a detector $P$ (i.e., InfoMap), which is more powerful than detector $D$ (i.e., modularity), and the output of $P$ places almost all negative ties between communities, one can confidently state that the negative ties are not informative for the detection task (``more powerful" qualitatively refers to a detector that finds more cohesive groups, and wrongly clusters distinct communities due to having more inter-connections). Furthermore, if one can find a detector $M$ that is more powerful  than $P$, the statement is still valid, since detector $M$ further splits the communities of $P$, rather than clustering them together.

Knowing that InfoMap is more powerful than Modularity in non-signed mode (see Refs. \cite{aldecoa2013exploring,lancichinetti2009community,kawamoto2014map,orman2012comparative}), we consider the objective function of detector $P$ as follows:
\begin{equation}
L(G,C)=\alpha L(G^+,C) - (1-\alpha) L(G^-,C)
\end{equation}
where $L(G,C)$ is the generalization of InfoMap's objective function. Note that there is still no exact formulation for $\alpha < 1$, nonetheless, we suppose it will be devised in the future, and will outperform modularity for $\alpha < 1$.

According to the results, for $\alpha = 1$, detector $P$ places almost all of negative ties between communities, and thus they have no contribution to $L(G^-,C)$, as it only punishes internal negative ties. In addition, we showed in Sec. \ref{subsec:info} that the remaining internal negative ties are incapable of ripping the unbalanced communities apart, mainly because they are supported by a large number of positive ties. Therefore, even if a general objective $L(G,C)$ is proposed, it cannot considerably improve upon $L(G^+,C)$  for $\alpha < 1$. Moreover, as we previously argued, this statement also holds for even more powerful detectors than $P$.

It could be concluded that, in the case of Epinions, information of negative ties is helpful for signed modularity, which is also the case for weaker methods like detector $D$.  However, by the use of nonsigned  InfoMap, which performs at least as well as modularity, together with the information analysis of internal negative ties, one can conclude that negative ties are not considerably informative for community detection in Slashdot and Epinions.
\section{Community-Community Interactions}
As discussed in Sec. \ref{sec:comcls}, one should let positively-related communities be separated from each other. This is the main goal of all community detection methods for nonsigned networks, which have been extended to be fit for signed networks. However, from the GSB perspective, this discrimination is not allowed and has been questioned by Doreian and Mrvar \cite{doreian2009partitioning}. As a result, they proposed RSB that allows positive relationships between two clusters. In particular, RSB was successfully applied on some small-scale networks that a complete scenario of their relations was known. However, this assertion, to the best of our knowledge, has not yet been examined on large-scale networks. Indeed, in this work we try to answer the question, \textit{``Are mediator triads frequent enough in signed networks, to be considered in clustering algorithms?"}.

First, we should provide a connection between the output of InfoMap and GB modeling. The GB model does not impose any restrictions on built-in structure of each cluster and leaves it to the algorithm to find a suitable type or to the researcher to pre-specify it based his or her knowledge \cite{doreian2005generalized}. However, for large-scale datasets like those analyzed in this work, this cannot be efficiently done mainly due to the lack of data about the history of individuals. This leaves us with only one option, to use the general assumption that social clusters are likely to be densely connected  \cite{newman2003social, newman2003structurefunc, backstrom2006group}. This assumption is a special case in GB modeling known as \textit{complete block}, which is used for detection of cohesive subgroups \cite{doreian2005generalized}. With this restriction, we are allowed to investigate the mediator clusters in social networks based on the output of InfoMap, and further probe the community-community relations.

Let us define some quantities to investigate community-community interactions quantitatively:
\begin{equation}
F(i, j)=\left\{\begin{matrix}
1 & \mbox{all inter-edges are positive}\\ 
0 & \mbox{o.w.}
\end{matrix}\right.,
\end{equation}
\begin{equation}
E(i, j)=\left\{\begin{matrix}
1 & \mbox{all inter-edges are negative}\\ 
0 & \mbox{o.w.}
\end{matrix}\right.,
\end{equation}
where (i, j) is a pair of communities. Less than 6\% (2\%) of community-community relations  are ignored due to partial negative-positive ties in Slashdot (Epinions).

As depicted in Table \ref{tab:basic_meso1}, more than 66\% (in Slashdot) and 79\% (In Epinions) of community-community relations are friendship, and on average, a balanced community is friends with around 69\% (in Slashdot) and 77\% (In Epinions) of its neighbor communities. On the other hand, less than 12\% of communities are mostly enemies with their neighbors. 

\begin{figure}
\centering
\includegraphics[width=\figratio\textwidth]{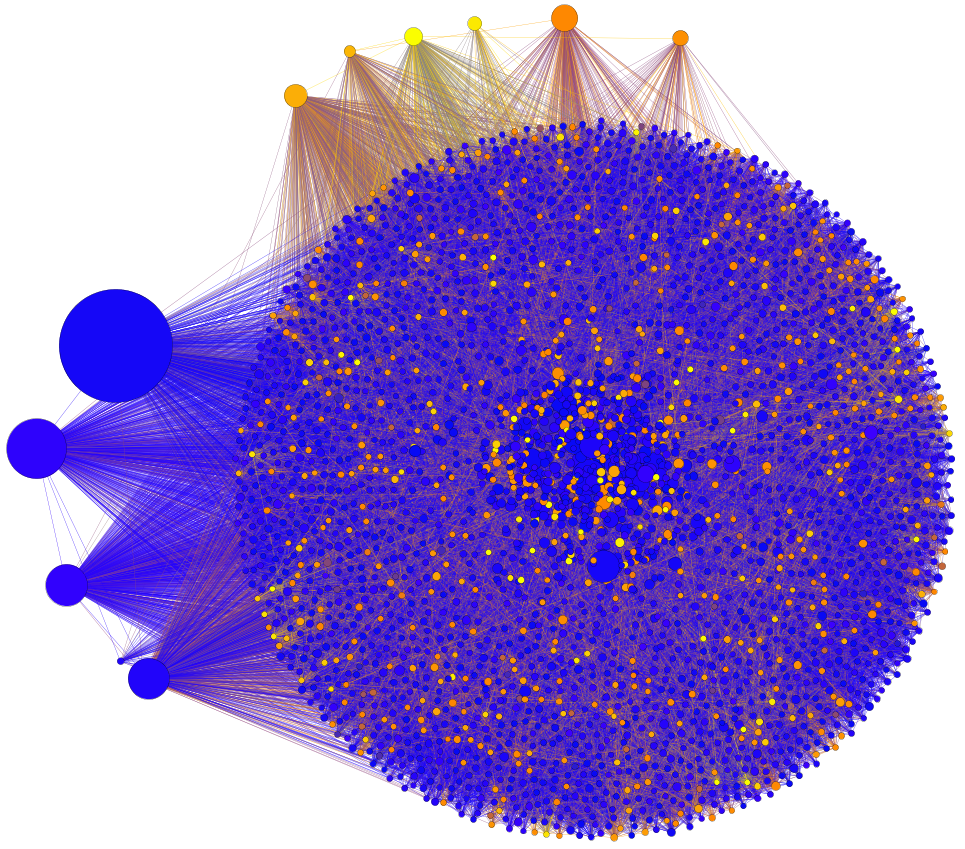} 
\caption{(Color) Constructed community network of Slashdot. All communities are internally balanced. The size of each community is proportional to the number of its members. Communities with higher ratio of negative relations are closer to yellow color (lighter gray). Some of the larger aggressive communities, which are the enemies of the majority of their neighbors, have been placed on the top of the network. These communities indicate a meaningful group of people that are allies, and troll the neighbor communities.}
\label{fig:slashdot}
\end{figure}

In Fig. \ref{fig:slashdot}, the Slashdot network of balanced communities is visualized using Gephi\footnote{Gephi is an open source software for visualizing large-scale graphs: \url{https://gephi.org/}}. It is worth mentioning that the work of Kunegis \textit{et al.} \cite{kunegis2009slashdot}  investigated the Slashdot trolls individually (users that are the enemy of most of their neighbors). In the mesoscopic level, our result in Fig. \ref{fig:slashdot} shows some of the major trolling communities unearthing a novel view of Slashdot. The $k$-clustering algorithms on social networks merely distinguish the aggressive groups. However, low amounts of these clusters indicates that clustering algorithms, even in optimal case, could miss detecting a considerable amount of distinguishable and positively-related groups by merging them into one another, similar to Fig. \ref{fig:ClusCom}(C). Therefore, clustering based on social balance hides a major part of the mesoscopic structure.
\begin{table}
\centering
\caption{Statistics of community networks. Each node represents a balanced community and each link is positive (negative) if $F(i,j)=1$ [$E(i,j)=1$]. For each community, friendship (enmity) is the percentage of positive (negative) degree to total degree. A community is friendly if its Friendship is larger than 80\%. A community is aggressive if its enmity is larger than 50\% (similar thresholds do not considerably change the results).}
\label{tab:basic_meso1}
  \centering
    \begin{tabular}{lcc}
    \hline\hline
    \toprule
          & Slashdot & Epinions \\
    \midrule
    Friendly-Relations & 66\%  & 79\% \\
    Enmity-Relations & 29\%  & 19\% \\
    Average no. Neighbors & 58    & 28 \\
    Average Friendship & 69\%  & 77\% \\
    Average Enmity & 28\%  & 22\% \\
    Friendly Communities & 31\%  & 58\% \\
    Aggressive Communities & 12\%  & 12\% \\
    \bottomrule
    \hline\hline
    \end{tabular}
\end{table}

\subsection{Mediator Triads should not be Overlooked}
\label{subsec:freqmed}
 The relaxation of Doreian and Mrvar could be justified only if the mediator triad appears frequently enough in the mesoscopic level of social relations. In particular, RSB does not claim that the frequency of mediator triads, relative to a null model, is either underrepresented or overrepresented. Nonetheless, it legitimizes the absolute presence of such relations. Therefore, if there is a considerable amount of such relations, even underrepresented, the RSB should be utilized instead of GSB.
 
 In he local level, as shown in Table \ref{tab:triad_freq}, the $++-$ triad is the only one that is considerably underrepresented, which is 11\% (in Slashdot) and 8\% (in Epinions) of triads in real networks against 41\% (in Slashdot) and 35\% (in Epinions) of triads in randomized counterparts. This observation is consistent with the previous findings in favor of GSB (excluding $k = 2$) \cite{leskovec2010signed, szell2010multirelational}. 

In the mesoscopic level, we conduct a similar experiment on the network of balanced communities. To this end, we consider each community as a node, and connect a pair of communities with $+1 (-1)$ edge, if they were friend (enemy). As summarized in Table \ref{tab:basic_meso2}, the community network has similar percentage of negative edges compared to the local level network. This makes the comparison of two levels more meaningful.

As depicted in Table \ref{tab:triad_freq}, the ratio of mediator triads is 34\% (in Slashdot) and 21\% (in Epinions), which is by far higher than 11\% (in Slashdot) and 8\% (in Epinions) in local level. As listed in the \textit{z-score} column, mediator triads are also underrepresented according to randomized networks. This means even if this type of triad is a less desirable relation between groups of individuals, nonetheless, this is a notable pattern among them. Hence, as suggested by Doreian and Mrvar,  ignoring the intermediary processes leads to merging or even splitting a considerable amount of mediators into hostile parties. In other words, if one has low amount of mediator triads close to that of a $k$-balanced network, ignoring the mediator triads does not conceal the true mesoscopic structure. However, observed frequencies suggest that although mesoscopic structures are driven away from mediator triads (according to corresponding z-scores), the considerable amount of these relations refrain us from simply ignoring them. Moreover,  explicitly established relations between users leave no room for the assumption that the mediator triads are mainly due to noise.

In conclusion, although the mediator triad is underrepresented in social networks, it should not be overlooked. In particular, the implication of GSB remains valid in the sense that social dynamics drive the relations away from the mediator triad. Nonetheless, the relaxation of Doreian and Mrvar is stil necessary to account for mediator triads, which are still surviving the social dynamics, as a remarkable aspect of social relations.
\begin{table}
\centering
\caption{Total number of triads and percentage of negative links in the original Slashdot and Epinions networks and the constructed community network. $M$ and $K$ stand for $10^6$ and $10^3$, respectively.}
\label{tab:basic_meso2}
\begin{tabular}{|c|c|c|c|c|}
	\hline
    \toprule
    \multirow{2}[4]{*}{} & \multicolumn{2}{c|}{Local} & \multicolumn{2}{c|}{Mesoscopic} \\ \cline{2-5}
    \midrule
          & no. Triads & Negative ties & no. Triads & Negative ties \\
    \hline
    Slashdot & 0.6M  & 23.60\% & 284K  & 30.40\% \\
    \hline
    Epinions & 4.8M  & 16.80\% & 35K   & 20.20\% \\
    \bottomrule
    \hline
    \end{tabular}
\end{table}

\begin{table*}
\centering
\caption{Percentage and z-scores of each triad type in the local and mesoscopic levels compared to $M_{100}(G)$. Statistics for the $++-$ triad, which is the only  unbalanced triad in GSB (excluding $k = 2$), are shown in bold.}
\label{tab:triad_freq}
\begin{tabular}{|c|c|c|c|c|c|c|c|c|c|c|c|c|c|}
	\hline
    \toprule
    \multirow{3}[6]{*}{} & \multicolumn{6}{c|}{Slashdot}                  & \multirow{7}[14]{*}{} & \multicolumn{6}{c|}{Epinions} \\ \cline{2-14}
    \midrule
          & \multicolumn{3}{c|}{Local} & \multicolumn{3}{c|}{Mesoscopic} &       & \multicolumn{3}{c|}{Local} & \multicolumn{3}{c|}{Mesoscopic} \\ \cline{2-14}
          & Real  & Random  & z-score & Real  & Random  & z-score &       & Real  & Random  & z-score & Real  & Random  & z-score \\
    \hline 		
    +++   & 73    & 44.6  & 90    & 38.2  & 33.7  & 16    &       & 82.6  & 57.6  & 149   & 66.3  & 50.7  & 24 \\
    \hline
    \textbf{++ -} & \textbf{11.2} & \textbf{41.3} & \textbf{-232} & \textbf{33.7} & \textbf{44.2} & \textbf{-105} &       & \textbf{8.3} & \textbf{34.9} & \textbf{-261} & \textbf{20.8} & \textbf{38.6} & \textbf{-45} \\
    \hline
    + - -   & 13.6  & 12.7  & 5     & 21.9  & 19.3  & 13    &       & 7.9   & 7     & 14    & 11    & 9.8   & 4 \\
    \hline
    - - -   & 2.1   & 1.3   & 23    & 6.2   & 2.8   & 58    &       & 1.1   & 0.5   & 93    & 1.9   & 0.8   & 17 \\
    \bottomrule
    \hline
    \end{tabular} 
\end{table*}

\section{Conclusion}
\label{sec:conc}
In this work, we investigated the mesoscopic level of online signed social networks. First, we observed that communities (extracted based on merely positive edges) in signed social networks are highly balanced. This indicates that negative edges mostly lie between dense positive clusters. Also, when negative edges lie inside the communities, they have either no or weak divisive power. In other words, negative edges do not have a significant effect on the community structure of signed networks, and it is mainly determined by positive relations. Furthermore, we showed that this salient characteristic is almost impossible to be created by randomly placed negative edges. This assertion is consistent with the previous studies both on the local level, where it was shown that the clustering coefficient of positive subgraph is much higher than that of negative subgraph \cite{leskovec2010signed}, and the global level, where it was demonstrated that social networks are highly balanced compared to sign-shuffled ones \cite{facchetti2011computing}. This role of negative ties partially explains why sign prediction models that are based on machine learning techniques can perform highly accurately, despite the fact that they utilize the information of merely adjacent nodes \cite{leskovec2010predicting, tang2012inferring}. Our second observation was that the $++- $ mediator triad between communities is underrepresented consistent with GSB; however, it is highly frequent compared to the triad of the same type between users. Hence, mediator triads cannot be simply ignored as they still survived the social dynamics and form a considerable portion of social relations. As a result, if one only tries to minimize $F_k(G,C)$ regardless of the mediator triads, many intermediary clusters are lost by merging or splitting them into hostile parties, and hence, major parts of the mesoscopic structure remain hidden. Consequently, the routes of RSB-based GB modeling and signed community detection seem to be more consistent with the structure of networks similar to Slashdot and Epinions.
\section{Future Works}
\label{sec:future}
There are some interesting issues that can be investigated in future works, including:
\begin{itemize} \itemsep4pt \parskip0pt \parsep0pt
\item[\textbullet]
In this work, we measured the informativeness of negative edges for each community separately. It is fruitful to have a procedure that measures this information in a network as a whole. Although signed modularity can do this work, its major shortcomings make it an unreliable measure for real-world networks \cite{fortunato2007resolution, good2010performance, lancichinetti2011limits}. Nonetheless, along with $F_k(G,C)$, it can be a baseline for future measures.
\item[\textbullet]
An improvement in accuracy of the link prediction is likely to be achieved by augmenting the (nontrivial) statistics of InfoMap communities into machine learning methods. Noting that a successful work has been carried out by extracting clusters [detected via minimizing $F_k(G,C)$] and further applying collaborative filtering methods \cite{javari2014cluster}.
\item[\textbullet]
We showed when negative ties lie between dense positive ties, their informativeness vanishes for the task of community detection. On the contrary, as demonstrated in Ref. \cite{leskovec2010predicting}, they are really useful for inferring hidden links due to this apparent pattern. Roughly, the less the usefulness of negative ties for community detection, the more their usefulness for link prediction. Thus, an interesting task would be a quantitative analysis of interplay between the information of negative ties in the local level (for link inference) and that of the mesoscopic level (for community detection).

\end{itemize}

\bibliography{refs}

\end{document}